\documentclass[a4paper,fleqn]{cas-sc}
\usepackage[numbers,sort&compress,square]{natbib}
\usepackage{multirow}
\usepackage{amsmath}
\usepackage{natbib}
\def\tsc#1{\csdef{#1}{\textsc{\lowercase{#1}}\xspace}}
\tsc{WGM}
\tsc{QE}
\tsc{EP}
\tsc{PMS}
\tsc{BEC}
\tsc{DE}
\begin{document}
\let\WriteBookmarks\relax
\def\floatpagepagefraction{1}
\def\textpagefraction{.001}
\shorttitle{Breast MRIs Normalization}
\shortauthors{Modanwal et~al.}

\title [mode = title]{Normalization of Breast MRIs using Cycle-Consistent Generative Adversarial Networks}                      
% \tnotemark[1,2]

% \tnotetext[1]{This document is the results of the research
%   project funded by the National Science Foundation.}

% \tnotetext[2]{The second title footnote which is a longer text matter
%   to fill through the whole text width and overflow into
%   another line in the footnotes area of the first page.}

\author[1]{Gourav~Modanwal}[type=editor,
                    %auid=000,bioid=1,
                    %prefix=Dr.,
                    %role=Researcher,
                    orcid=0000-0001-6215-4687]
%\cormark[1]
\ead{gourav.modanwal09@gmail.com}
%\ead[url]{www.cvr.cc, cvr@sayahna.org}

%\credit{Conceptualization of this study, Methodology, Software}

\address[1]{Department of Radiology, Duke University, Durham, NC}

\author[2]{Adithya~Vellal}
%\ead{adithya.vellal@duke.edu }

\author[1]{Maciej A. Mazurowski}
%[%   role=Co-ordinator,   suffix=Jr,   ]
%\ead{maciej.mazurowski@duke.edu}
%\ead[URL]{www.sayahna.org}

%\credit{Data curation, Writing - Original draft preparation}

\address[2]{Department of Computer Science, Duke University, Durham, NC, USA}

%\author%
% [1,3]
% {Rishi T.}
% \cormark[2]
% \fnmark[1,3]
% \ead{rishi@stmdocs.in}
% \ead[URL]{www.stmdocs.in}
% \address[3]{STM Document Engineering Pvt Ltd., Mepukada,
%     Malayinkil, Trivandrum 695571, India}
%\cortext[cor2]{Principal corresponding author}
%\fntext[fn1]{This is the first author footnote. but is common to third
% \fntext[fn2]{Another author footnote, this is a very long footnote and
%   it should be a really long footnote. But this footnote is not yet
%   sufficiently long enough to make two lines of footnote text.}
%\nonumnote{This note has no numbers. In this work we demonstrate $a_b$
% the formation Y\_1 of a new type of polariton on the interface
% between a cuprous oxide slab and a polystyrene micro-sphere placed
% on the slab.
% }
\begin{abstract}
\noindent{Objectives: Dynamic Contrast Enhanced-Magnetic Resonance Imaging (DCE-MRI) is widely used to complement ultrasound examinations and x-ray mammography for early detection and diagnosis of breast cancer. However, images generated by various MRI scanners (e.g., GE Healthcare, and Siemens) differ both in intensity and noise distribution, preventing algorithms trained on MRIs from one scanner to generalize to data from other scanners. In this work, we propose a method to solve this problem by normalizing images between various scanners.}

\noindent{Methods: MRI normalization is challenging because it requires normalizing intensity values and mapping noise distributions between scanners. We utilize a cycle-consistent generative adversarial network to learn a bidirectional mapping and perform normalization between MRIs produced by GE Healthcare and Siemens scanners in an unpaired setting. Initial experiments demonstrate that the traditional CycleGAN architecture struggles to preserve the anatomical structures of the breast during normalization. Thus, we propose two technical innovations in order to preserve both the shape of the breast as well as the tissue structures within the breast. First, we incorporate mutual information loss during training in order to ensure anatomical consistency. Second, we propose a modified discriminator architecture that utilizes a smaller field-of-view to ensure the preservation of finer details in the breast tissue.}

\noindent{Results: Quantitative and qualitative evaluations show that the second innovation consistently preserves the breast shape and tissue structures while also performing the proper intensity normalization and noise distribution mapping.}

\noindent{Conclusion: Our results demonstrate that the proposed model can successfully learn a bidirectional mapping and perform normalization between MRIs produced by different vendors, potentially enabling improved diagnosis and detection of breast cancer. \textit{All the data used in this study are publicly available at \url{https://wiki.cancerimagingarchive.net/pages/viewpage.action?pageId=70226903}.}
}
\end{abstract}

% \includegraphics[width=6.50in]{Graphical_Abstract.pdf}

% \begin{highlights}
% %\begin{itemize}
% \item We present a method for MRI normalization that performs unpaired bidirectional normalization between DCE-MRIs produced by different scanner models.
% \item We investigate the challenges of standard CycleGAN for the normalization of MRI and propose two technical solutions to overcome it.
%             \begin{itemize}
   
%             \item Incorporation of a mutual information loss with standard CycleGAN.
%             \item Modified discriminator capable of preserving breast shape and dense tissues.
            
%             \end{itemize}
% \item  Quantitative and qualitative evaluations show that modified discriminator consistently preserves breast shape and tissue structures while also performing proper intensity normalization and noise distribution mapping.
% %\end{itemize}
% \end{highlights}


\maketitle

\section{Introduction}

Breast cancer is one of the leading causes of death among women around the globe \cite{ACS_report}. Dynamic contrast-enhanced magnetic resonance imaging (DCE-MRI) is widely used to complement mammography and ultrasound when evaluating breast cancer, particularly when assessing the extent of cancer before surgery \cite{reig2020machine}. In some high-risk cases, it is also used for screening.

A significant challenge related to the use of DCE-MRI is the lack of standardized imaging protocols \cite{sachs2017ct,sharma2020standardizing}. Different MRI scanners use different parameters, which previous research \cite{saha2017effects} has shown to drastically alter image appearance, quality, as well as the radiomics analysis. When the same patient is imaged using a different scanner or even the same scanner with different scanner parameters, the produced MR images may vary significantly \cite{nyul2000new,rizzo2018radiomics}. The inconsistencies present in the radio-frequency (RF) coil produce intensity variations in the underlying tissue across the scanned image \cite{roy2011intensity}. Additionally, varying scanner parameters alter the noise distribution of the images. An illustration of the difference in intensity and noise distribution between images obtained from two different MRI scanner manufacturers (GE Healthcare and Siemens) is shown in Fig.~\ref{fig:sample images}. 

The high degree of inter-scanner variation proves to be a significant obstacle to the effective usage of DCE-MRI. In the context of radiomics, where a multitude of features are extracted from images for further processing, the features from different modalities may turn out to be incomparable, thus rendering them useless for classification and prediction. The impact of scanner parameters on breast MRI radiomic features is demonstrated by \cite{saha2017effects}. Variability in images has been shown to have an impact on the training of deep learning as well \cite{albadawy2018deep}. Algorithms trained on images from one scanner may not perform well on exams at a different institution that was acquired using a different scanner \cite{maartensson2020reliability}. Finally, the inconsistency between images from different scanners may affect the outcome of computer-aided diagnosis. The ability to translate and normalize between images acquired by different vendors with varying parameters of scanner would have tremendous positive consequences. It would enable quantitative comparison of image features across various institutions. It would also improve generalization as deep models trained on one dataset could still perform inference on new datasets generated by different scanners.

In order to address this issue, we frame the problem of normalization between images generated by different MRI scanners as an application of unpaired image-to-image translation. Most of the literature in the domain of MRI pre-processing has focused on normalizing intensities but does not account for noise patterns. To our knowledge, no one has yet proposed a method for MRI vendor normalization. This process is challenging because it requires both normalizing the intensity and learn the mapping between the noise patterns. In this work, we present a vendor normalization method that attempts to perform intensity normalization as well as noise distribution mapping between MRIs obtained from different scanners. The significant contributions of this work can be summarized as follows:

\begin{itemize}
\item We present a method for MRI vendor normalization that performs unpaired bidirectional normalization between DCE-MRIs produced by different scanner models.
	\item We investigate the challenges of the standard CycleGAN approach for normalization of medical images, primarily the difficulty in maintaining the breast shape and structures within the breast between the original image and the normalized image. Then, we propose and evaluate two technical solutions to this issue, as described below.
        \begin{itemize}

        	\item We propose the incorporation of a mutual information loss with the standard CycleGAN architecture in order to ensure that the breast shape and tissue structures within the breast is maintained.
        \item We propose a modified discriminator capable of preserving the breast shape as well as the dense tissues and evaluate the effect of changing the field-of-view on the performance.
        
        \end{itemize}

\begin{figure}[!t]
\centering
\includegraphics[width=3.57in]{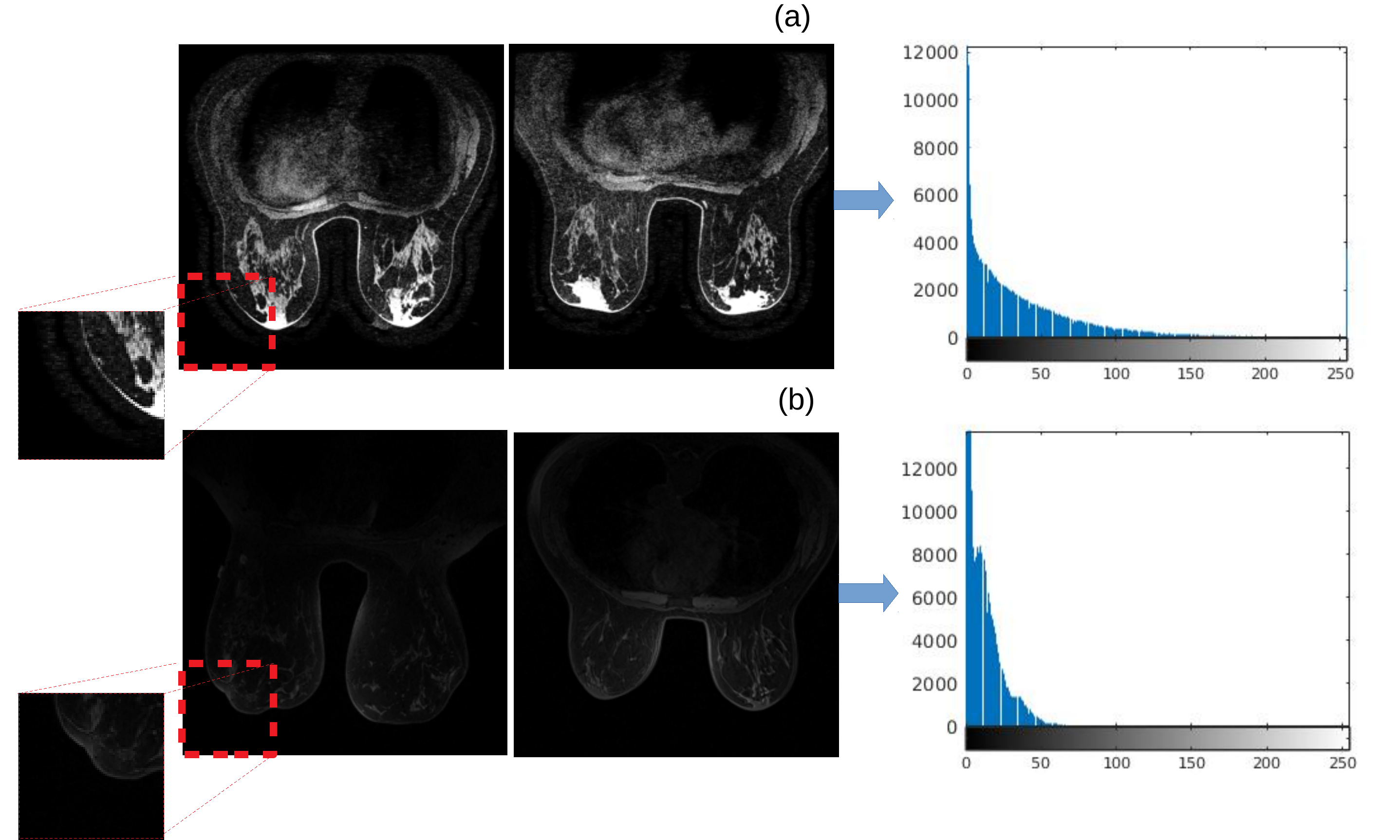}
\caption{Example of images from two scanner displaying differences in intensity and noise distribution (a) GE Healthcare (b) Siemens.}
\label{fig:sample images}
\end{figure}

\end{itemize}

We further present and compare the performance of the proposed vendor normalization methods using both quantitative and qualitative approaches.
We also highlight how the proposed work can potentially enable the synthesis of larger and richer datasets that mitigate issues related to class imbalance.

The remainder of the article is organized as follows. Section~\ref{Sec:Rel_work} describes the related work.
Section~\ref{Sec:Dataset} presents details about the dataset, and the proposed methods are detailed in 
Section~\ref{Sec:Methods}. Information about training is furnished in Section~\ref{Sec:Training} and  Section~\ref{Sec:Evaluation} presents the metrics for the evaluation of the proposed method. Section~\ref{Sec:Result} reports the experimental results and discussions. Finally, Section~\ref{Sec:Conclusion} concludes with a summary.

\section{Related Work}
\label{Sec:Rel_work}

Unlike other imaging modalities, MRIs span a wide, non-linear spectrum of raw intensity values. They lack uniformity and often exhibit high variance between subjects. Even within a single subject, intensity variations of 10-40\% have been observed \cite{simko2019generalized}. This heterogeneity makes it difficult to effectively train robust medical image analysis algorithms on MRIs \cite{pooch2019can}.

In response, various statistical approaches have been proposed for MRI intensity normalization. These include histogram equalization \cite{nyul2000new,sun2015histogram}, intensity scaling based on regions of interest \cite{collewet2004influence} and landmarks \cite{madabhushi2006new}. However, histogram-based methods rely on discrete approximations of intensity distributions, leading to high levels of inexactness \cite{roy2011compressed}. Meanwhile, obtaining a high level of accuracy with landmark-based algorithms requires obtaining multiple landmarks from various tissue types in the image. Designing algorithms to perform this landmark selection task is difficult and time-consuming \cite{madabhushi2006new}. Another limitation of many MRI normalization methods \cite{shinohara2014statistical,zhang2018automatic} is that they require auxiliary inputs such as segmentation masks. This adds an intrinsic reliance on the models that perform these preprocessing tasks. Alternatively, some techniques \cite{fischl2004sequence} attempt to leverage the physics of MR acquisition in order to develop intensity invariant segmentation algorithms. However, using this type of approach requires integrating explicit physics-based embedding into the segmentation algorithm, thus limiting this system's ability to generalize to other downstream tasks.

Additionally, some of the methods discussed above \cite{nyul2000new,collewet2004influence,madabhushi2006new} attempt to perform intensity transformation between two fixed imaging settings. That is, they make the assumption that the intensity relationship of the tissues is constant between the target group and the reference group, which is not always true \cite{gao2019universal}. If the intensity standardization needs to be done for images coming from multiple centers, multiple transforming models need to be established. Resultantly, these methods do not have the ability to process new images that are not from an MR image group that has already been included in their training data. This severely limits its usability. Work \cite{van2020harmonization} presents a model‐based method for harmonization between patients scanned with differences in imaging parameters.

%%%%%%%%%%%%%%%  % Go through section 3.2 of the article for more info.
Recently, GANs have been used in a variety of applications to the domain of medical imaging. An excellent review of GANs' recent applications to the medical domain was presented by \cite{yi2019generative_review,tschuchnig2020generative,morrison2021generative}. Most of the previous work \cite{cronin2020using,nie2018medical,jin2019deep,armanious2019unsupervised} has focused on using GANs for multimodal translation that in turn, improved diagnosis across several modalities (e.g., ultrasound, PET, CT, and MRI). Additionally, GANs have successfully generated synthetic images \cite{van2019strategies,dar2019image,sandfort2019data,shi2020knowledge,sun2020adversarial,shin2021style} to augment training datasets for algorithms that perform downstream tasks\textemdash diagnosis, prognosis, segmentation, and registration. Lately, GANs have also been used for normalizing MRIs across different scanners. Work by \cite{gao2019universal} proposed standardization method for brain MRIs using GANs with a weak paired data strategy with focus on intensity normalization only. Another work \cite{dar2019image} applies CycleGAN to brain MRI normalization. However, CycleGAN has an intrinsic ambiguity with respect to geometric transformations \cite{zhang2018translating}. More specifically, since the anatomical structure of the images in a set of patient data is highly variable, CycleGAN is unlikely to utilize anatomical features in order to determine the realness of an image. However, algorithms for multi-modal translation and synthesis of medical images should ensure shape consistency, as these anatomical structures are crucial information for computer-based cancer detection algorithms.

One of the research work by \cite{zhang2018translating} has tried to solve the problem by adding an extra penalty based on a segmentation mask generated from the CycleGAN output. However, this requires a ground-truth annotation of the dense tissue regions in the breast, which is not available in a typical use case for breast MRI normalization. 
Meanwhile, \cite{wang2018unsupervised} introduce deformable convolutional layers and novel cycle-consistency losses. Other papers \cite{cronin2020using,nie2018medical,jin2019deep,armanious2019unsupervised} utilize CycleGAN for translation between various modalities (e.g., ultrasound PET, CT and MRI) and reports many of the same issues discussed above. In this article, we present a fully unpaired algorithm for image normalization using CycleGAN. We propose and evaluate two technical solutions in order to effectively preserve the breast shape and tissue structures within the breast MRI.

\section{Dataset and Pre-processing}
\label{Sec:Dataset}
In this study, we utilize Duke-Breast-Cancer-MRI data\footnote{All images used in this study are publicly available at \url{https://wiki.cancerimagingarchive.net/pages/viewpage.action?pageId=70226903}} obtained from GE Healthcare (GE) and Siemens (SE) scanners (1.5 T) in the axial plane. Our dataset consists of 124 subjects: 77 imaged with a GE Healthcare scanner, and the remaining 47 with a Siemens scanner. Details about the distribution of patients \& scanner parameters across different manufacturer is presented in Table~\ref{tab:Scanner_Para}. Each MR volume contains more than 160 2D axial image slices. The top 1\% of pixel values in the entire dataset are truncated at 255, and the remaining intensities are linearly scaled to the 0-255 pixel range. The dataset are randomly divided into train, validation, and test set respectively at the patient level. We only use slices from the middle 50\% of each patient volume throughout our experiment. Details regarding the number of slices used for training, testing, and validation are given in Table~\ref{tab:no_of_img}.

\begin{table}[!b]
\begin{center}
\caption{Distribution of patients \& typical scanner parameters across different manufacturer.}
\label{tab:Scanner_Para}
\begin{tabular}{llllllll} \hline\hline
Vendor/Manufacturer                                                                           & Scanner model    & Acquisition matrix            & Echo time & Repetition time &   \# Subjects \\ \hline \hline
\multirow{3}{*}{\begin{tabular}[c]{@{}l@{}}GE Healthcare\end{tabular}} &{Signa HDx}                 & $340 \times 340$ & 2.3 ms    & 5 ms            &       43        \\ \cline{2-6}

                                                                                          &  \multirow{2}{*}{Signa HDxt}               & $340 \times 340$ & 2.4 ms    & 5.5 ms          &     33          \\
                                                                                          &         & $384 \times 360$ & 2.4 ms    & 5.5 ms            &       1        \\ \cline{2-6}

\hline
\multirow{2}{*}{Siemens }          & \multirow{2}{*}{MAGNETOM Avanto}         & $448 \times 448$  & 1.4 ms    & 4.1 ms          &       30  \\
&        & $320 \times 320$ & 1.4 ms    & 4.1 ms            &       17
%                                                                                               & MAGNETOM TrioTim          & $448 \times 448$ & 1.4 ms    & 4.1 ms          &             
\\ \hline \hline
\end{tabular}
\end{center}
\end{table}

\begin{table}[b]
\begin{center}
\caption{Details about training, test and validation dataset.}
\label{tab:no_of_img}
\begin{tabular}{lll} \hline \hline
      & GE Healthcare (GE) & Siemens (SE) \\ \hline \hline
Train Set & 5045          & 2776    \\ \hline
Test Set  & 1563          &  843    \\ \hline
Validation Set & 173          & 93    \\ \hline \hline
\end{tabular}
\end{center}
\end{table}

\section{Methods}

\label{Sec:Methods}
In this section, we present various frameworks to perform normalization between MRI images acquired by GE Healthcare (GE) and Siemens (SE) scanners.

\subsection{CycleGAN}

\begin{figure*} 
\centering

\includegraphics[height=6.41cm]{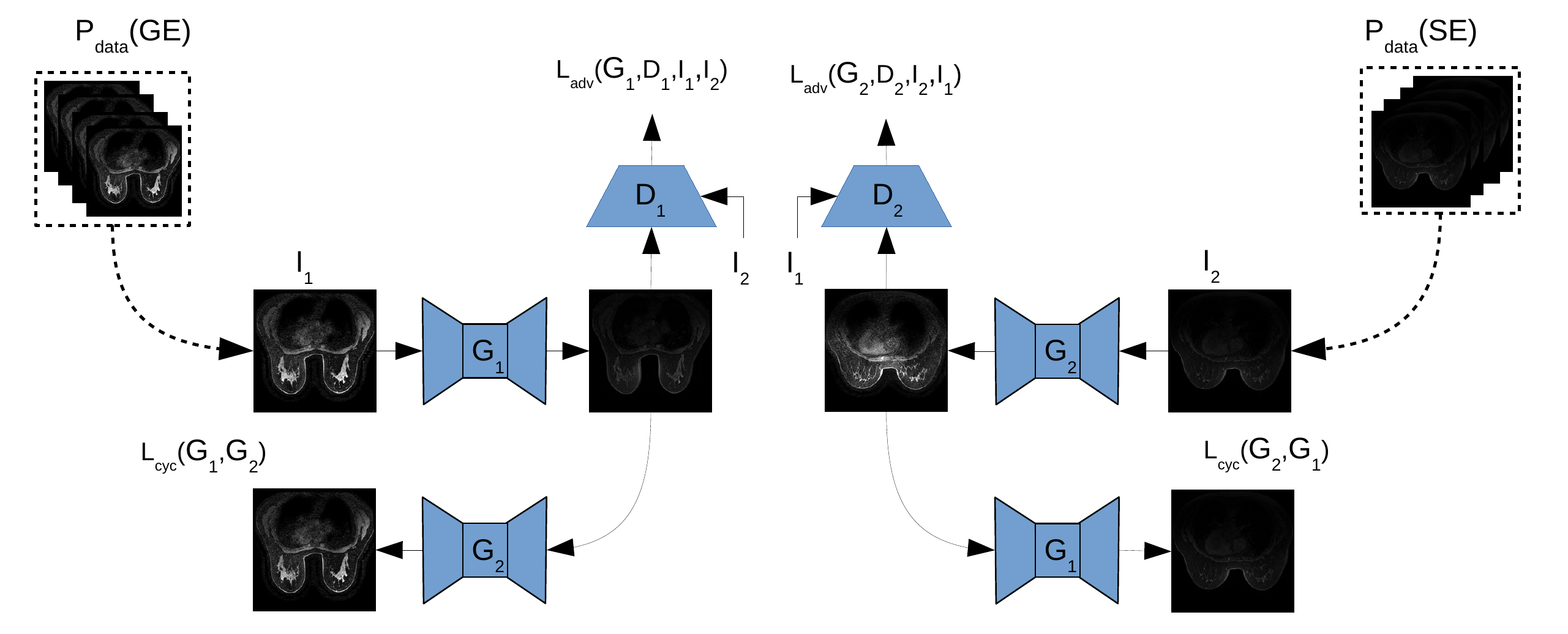}
\caption{ CycleGAN network configuration for breast MRI normalization with two generators $G_1$: normalizes $P_{data}(GE)$ $\rightarrow$ $P_{data}(SE)$ and $G_2$ normalizes $P_{data}(SE)$ $\rightarrow$ $P_{data}(GE)$, and associated adversarial discriminators $D_1$, $D_2$ . $I_1$ and $I_2$ are the unpaired training samples from $P_{data}(GE)$ and $P_{data}(SE)$, respectively.}
\label{fig:CycleGAN}
\end{figure*}

We utilize the CycleGAN \cite{Zhu_2017_Cyclegan}\textemdash a bidirectional image-to-image translation method\textemdash for the normalization between the GE and SE MRIs. It consists of two generators $(G_1,G_2)$ and two discriminators $(D_1,D_2)$. Each generator has a corresponding discriminator, and they are trained in an adversarial setting in which the two networks compete against each other to fool their counterparts. Fig.~\ref{fig:CycleGAN} illustrates the CycleGAN network configuration where, $I_1$ and $I_2$ are training samples from $P_{data}(GE)$ and $P_{data}(SE)$, respectively. Generator $G_1$ normalizes from $GE\rightarrow SE$ while $G_2$ normalizes from $SE\rightarrow GE$.

The discriminator network $D_1$ discriminates between the images generated by the generator $G_1(I_1)$ and the target image $I_2$ while generator $G_1$ tries to improve the quality of the transformed image so that it can fool the discriminator. Similarly, $D_2$ discriminates between images generated by $G_2(I_2)$ and the target image $I_1$, while $G_2$ tries to transform $I_2$ effectively enough to fool $D_2$.
The above task is formulated as a min-max optimization problem.

\subsubsection{Network Architectures}

The architecture for the generators is adapted from \cite{JohnsonAL16}. The generator consists of an encoder, transformer, and decoder. The encoder uses convolutional down-sampling to shrink the size of the input representation and increase the number of channels. It is followed by a transformation block which retains the size of representation using residual convolution blocks. Finally, a decoder block is used which upsamples the size of representation using deconvolution.

The discriminator network uses a classical PatchGAN \cite{pi2pix_patchgan}. It is a fully convolutional neural network that processes overlapping patches of the input image instead of the entire input image. The output of the discriminator is a matrix of binary classifications of whether each patch is real or fake. A standard PatchGAN has a field of view (FOV), or patch size, of $70 \times 70$. Our experiments with discriminator architectures with varying FOV are detailed in Section~\ref{Sec:Result}.

\subsubsection{Losses}

The objective function contains two loss terms: adversarial loss $(L_{adv})$ and cyclic loss $(L_{cyc})$. The adversarial loss \cite{goodfellow2014generative} ensures that the generated images belong to the data distribution of the target domain. The adversarial loss is formulated as below:

\begin{equation}
\begin{aligned}
\label{eq:eq:adv1}
L_{adv}(G_1,D_1,I_1,I_2)&=\mathbb{E}_{I_2\sim P_{data}(SE)}\left [ (D_1(I_2)-1)^{2} \right] 
+ \mathbb{E}_{I_1\sim P_{data}(GE)} \left [ (D_1(G_1(I_1)))^{2} \right]
\end{aligned}
\end{equation}

\begin{equation}
\begin{aligned}
\label{eq:eq:adv2}
L_{adv}(G_2,D_2,I_2,I_1)&=\mathbb{E}_{I_1\sim P_{data}(GE)}\left [ (D_2(I_1)-1)^{2} \right]
+\mathbb{E}_{I_2\sim P_{data}(SE)} \left [ (D_2(G_2(I_2)))^{2} \right]
\end{aligned}
\end{equation}

The generator tries to minimize the above adversarial loss, and the discriminator tries to maximize it. However, the adversarial loss alone is not sufficient enough to produce good target images. The adversarial loss will enforce the transformed output to be of the appropriate domain, but will not enforce the input and output to be recognizably the same. Thus an additional cycle-consistency loss is added to the overall objective. The cycle-consistency loss ensures that the translated image looks like the input image by enforcing $G_1$ and $G_2$ to be inverses of each other i.e. $(G_2(G_1(I_1)) \approx I_1$ and $(G_1(G_2(I_2)) \approx I_2$.

%\begin{equation}
\begin{equation}
\begin{aligned}
\label{eq:cyc}
L_{cyc}(G_1,G_2)=\mathbb{E}_{I_1\sim P_{data}(GE)}\left [ \left \| G_2(G_1(I_{1}))-I_1 \right \|_1  \right ] 
\end{aligned}
\end{equation}
% \end{equation}

\begin{equation}
\label{eq:cyc1}
L_{cyc}(G_2,G_1)=  \mathbb{E}_{I_2\sim P_{data}(SE)} \left [ \left \| G_1(G_2(I_{2}))-I_2   \right \|_1  \right ]
\end{equation}
% \end{equation}

The overall objective is given as below where $\lambda_{cyc}$ is the weighting factor for cycle-consistency loss.

\begin{equation}
\begin{aligned}
\label{Eq:Full_loss}
L(G_1,G_2,D_1,D_2)&=L_{adv}(G_1,D_1,I_1,I_2)  + L_{adv}(G_2,D_2,I_2,I_1) +\lambda_{cyc} * (L_{cyc}(G_1,G_2) \ +  L_{cyc}(G_2,G_1))
\end{aligned}
\end{equation}

\begin{figure} [!t]
\begin{center}
\begin{tabular}{c}
\includegraphics[height=5.5cm]{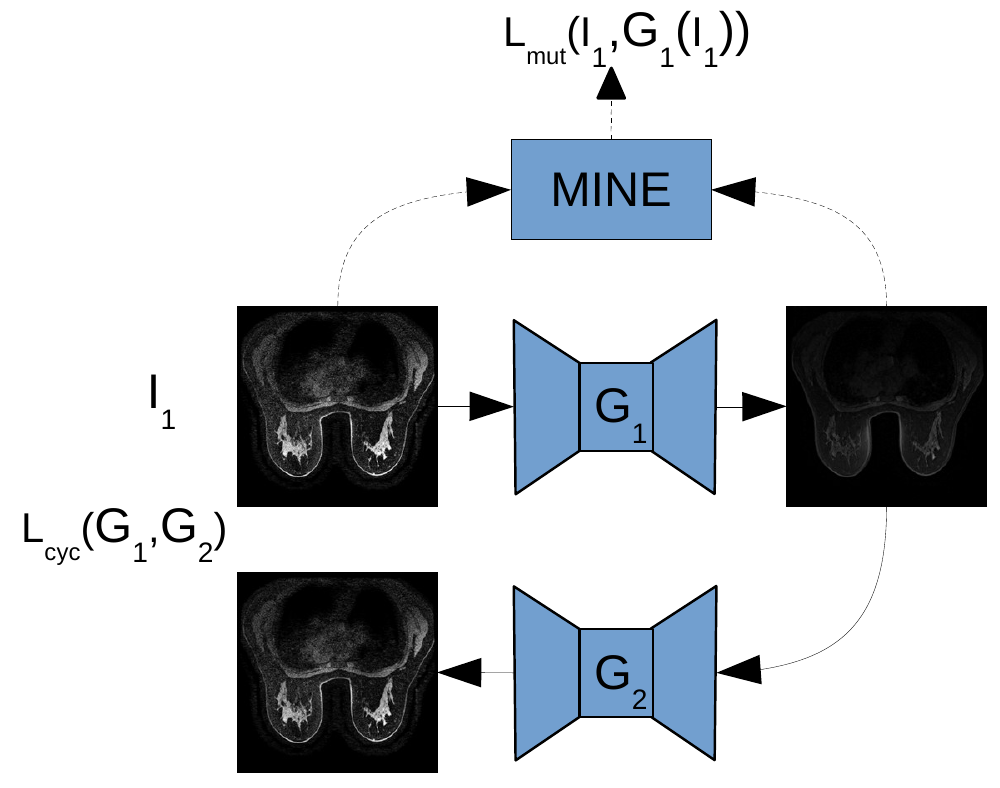}
\end{tabular}
\end{center}
\caption{An illustration of the proposed mutual information loss.}
\label{fig:MI_NW}
\end{figure}

\subsection{CycleGAN with Mutual Information}

The standard CycleGAN architecture detailed above, when used for normalization between GE and SE breast MRIs, may produce results that are unable to preserve the breast shape and tissue characteristics. In order to preserve the breast shape and tissue characteristics, we propose to utilize mutual information maximization between the real images and the generated images, as shown in Fig.~\ref{fig:MI_NW}. Our rationale is that while the intensity and texture of the image may change, high mutual information will indicate that the shape of the breast and the structure of dense tissue remained the same, which is desired in our application.

In practice, estimation of mutual information in images is challenging as we only have access to samples rather than the underlying distributions \cite{Poole2019OnVB_MI,mcallester2018formal}. Additionally, previous sample-based estimators are brittle and do not scale well to higher dimensions \cite{saxe2018information}. Recently, Mutual Information Neural Estimation (MINE) \cite{Belghazi2018MINE} was introduced to approximate the mutual information using observed samples even when the true distribution is unknown. Their approach also scales to higher dimensions. Hence, we adopt their method to estimate and maximize mutual information and utilize mutual information as a loss along with adversarial and cycle-consistency loss.

The mutual information is equivalent to the Kullback-Leibler (KL) divergence between the joint distribution, $P(X,Z)$, and the product of the marginal distributions $P(X)$ and $P(Z)$, as expressed below
\begin{equation}
\begin{aligned}
\label{Eq:KL_Diver_MI}
I(X,Z)=D_{KL}(P_{XZ}\left | \right |P_X \otimes P_Z)
\end{aligned}
\end{equation}
where $D_{KL}$ is defined as,
\begin{equation}
\begin{aligned}
\label{Eq:KL_Diver}
D_{KL}(P\left | \right |Q):= \mathbb{E_P}\left [log \frac{\partial P}{\partial Q}  \right ]
\end{aligned}
\end{equation}

It uses the Donsker\textendash Varadhan (DV) representation \cite{donsker1983asymptotic} of KL divergence, which leads to the following definition of approximate mutual information:
\begin{equation}
\begin{aligned}
\label{Eq:MINE_DV}
I_{\phi}(X, Z)=\sup_{\theta \in \phi }\left [\mathbb{E_{P_{XZ}}}[T_\theta ]-log(\mathbb{E_{P_X \otimes P_Z}}[e^{T_\theta}])  \right ]
\end{aligned}
\end{equation}

The approximate mutual information $I_{\phi}(X, Z)$ is obtained by maximizing the lower bound of the objective function shown in eq.~\ref{Eq:MINE_DV}. The maximization is achieved by using a neural network $(T_\theta)$ with parameters $\theta$. The neural network $(T_\theta)$ is optimized using gradient descent to characterise a family of functions which ultimately maximizes the lower bound of the above objective.

To enforce and preserve breast shape and tissue characteristics, we propose to include mutual information as a loss in the overall objective as specified below.

\begin{equation}
\begin{aligned}
\label{Eq:Full_loss_mut}
L(G_1,G_2,D_1,D_2)&=L_{adv}(G_1,D_1,I_1,I_2)\ + L_{adv}(G_2,D_2,I_2,I_1) 
+ \lambda_{cyc} * (L_{cyc}(G_1,G_2) +  L_{cyc}(G_2,G_1))  \\ 
&- \lambda_{mut} * (L_{mut}(I_1,G_1(I_1)) + L_{mut}(I_2,G_2(I_2)))
\end{aligned}
\end{equation}

where $\lambda_{mut}$ is the weight factor for mutual information loss.

\begin{figure} [t]
\begin{center}
\begin{tabular}{c}
\includegraphics[height=3.715cm]{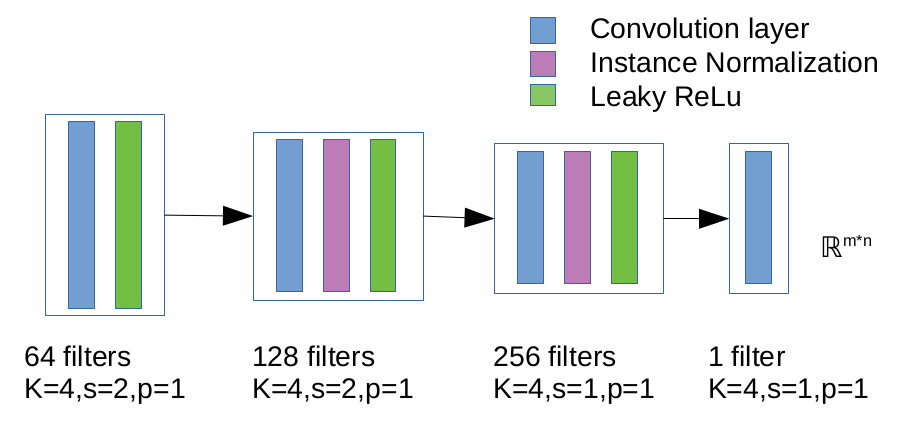}
\end{tabular}
\end{center}
\caption{Proposed discriminator architecture.}
\label{fig:Disc}
\end{figure}

\subsection{CycleGAN with modified discriminator}
\label{Sec:Disc}
We also modify our discriminator to test the effects of varying fields of view (FOV)\textemdash the size of the input pixel window that contributes to a single pixel in the output map. As suggested in \cite{pi2pix_patchgan}, we focus on smaller FOV to encourage the transformation learned by the generator to maintain sharp, high-frequency detail which is required in order to adequately preserve both the overall structure of the breast and the structure of the dense tissue regions inside the breast. Preliminary results were presented in \cite{modanwal2020mri}. Our experiments demonstrate that a ($34\times 34$) FOV discriminator architecture (shown in Fig.~\ref{fig:Disc}) is better at preserving morphological features of the breast tissue in comparison to the original ($70\times 70$) FOV. Quantitative analysis is presented in Table~\ref{Tab:FOV_Val}. Further details about the various discriminator architectures corresponding to different FOV are presented in appendix~\ref{appendix:Disc_Arc}.

\subsection{CycleGAN with modified discriminator + Mutual Information}
We also test the proposed Mutual Information loss in conjunction with the discriminator modification highlighted above.

\section{Training}
\label{Sec:Training}

We optimize the network using mean squared error (MSE) instead of cross-entropy, as suggested in \cite{Mao_2017_ICCV}. As a result, training becomes more stable, and higher quality images are produced. Additionally, to prevent the model from oscillation, the discriminator is fed a history of the 50 most recently generated images rather than solely the most recently generated image. Adam optimizer with the parameters $lr=0.002$, $\beta_1$ = 0.5, and $\beta_2$ = 0.999 is used to train the network weights.

With the addition of mutual information loss, the proposed framework has additional parameters to optimize. During experiments, we found that normalization quality is susceptible to these parameters. We optimized CycleGAN for different values of $\lambda_{cyc}$ (please refer Table \ref{Tab:Hyper_Cyclic_Loss}). The model with an optimal $\lambda_{cyc}$ (Std. CycleGAN) was used as a baseline for comparison. We then experimented for optimal $\lambda_{mut}$ (please refer Table \ref{Tab:Hyper_Mut_Loss}) and compared it with baseline model. The value of $\lambda_{cyc}$ and $\lambda_{mut}$ used in the experiments are 5.0 and 0.5 respectively.

%When $\lambda_{mut}$ is extremely small, mutual information loss will not take effect, and the result will be quite similar to baseline CycleGAN while a larger value of $\lambda_{mut}$ will emphasize more on mutual information loss and it will force the model to maximize mutual information. It will grow the shape of the breast and will make the result even worse. (Can be seen as reduced dice coff. with $\lambda_{mut}$ =1.0)

%Due to the time limit, we are only able to try a limited combination. Experimenting for better parameters is definitely an important and future work direction.

%\setcounter{figure}{4}
\begin{figure}[b]
\centering
\includegraphics[width=.9\linewidth]{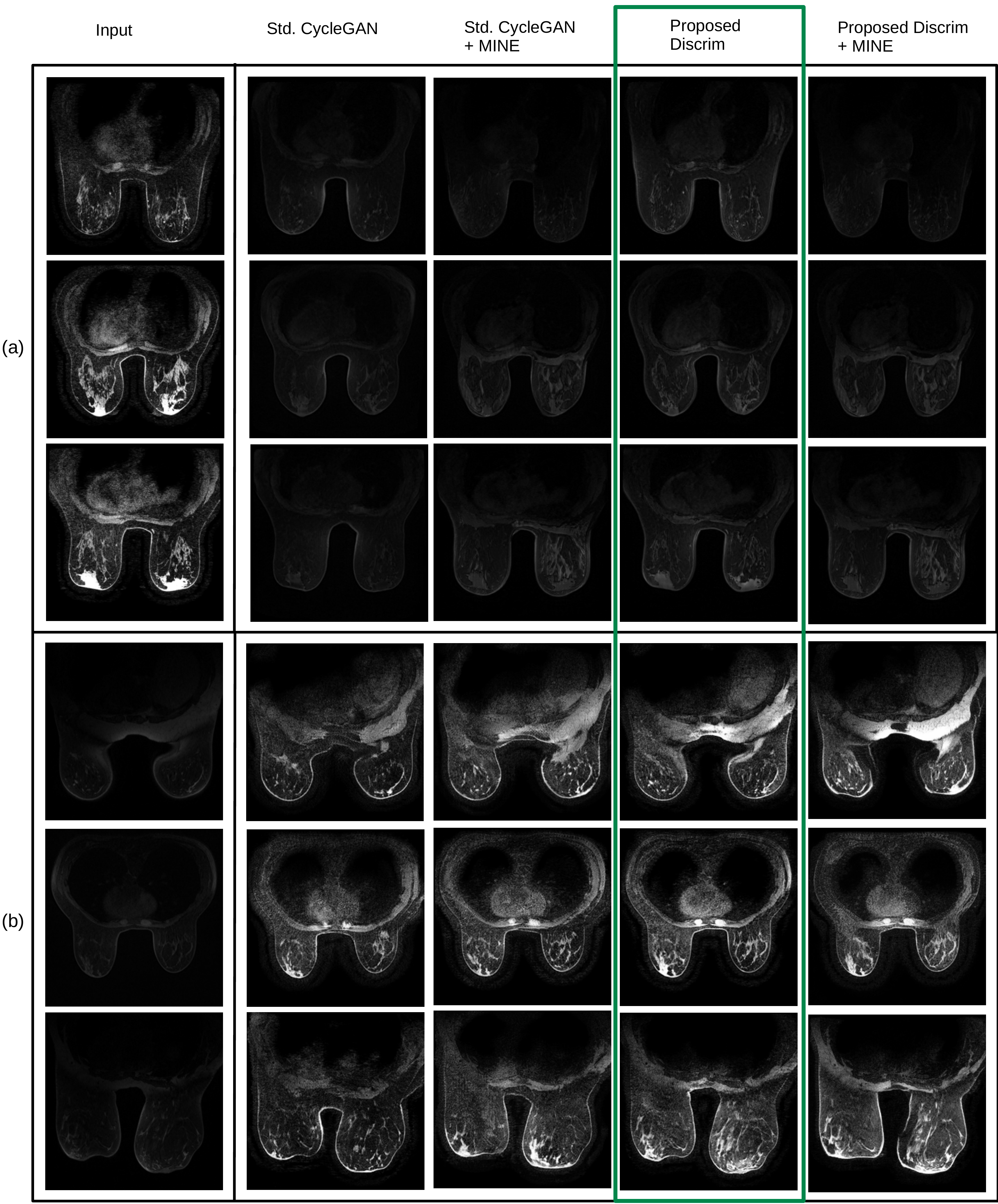}
\caption{Representative results of the proposed image normalization (a) GE Healthcare to Siemens (b) Siemens to GE Healthcare.}
\label{fig:Qual_Result}
\end{figure}

\section{Evaluation Metrics}
\label{Sec:Evaluation}

Quantitative evaluation of the normalized images is difficult in the case of unpaired images \cite{Zhu_2017_Cyclegan} as there is no standard/universal metric for assessing accuracy \cite{borji2019_metric_pros}. Hence, evaluating the quality of synthesized images is an open and challenging problem for which metrics vary depending on the specific needs of the application. Most previously published work relies either on the visual examination of the transformed images by human subjects or some application-specific metrics. Visual evaluation of the transformed image is still the most common and intuitive method for determining the quality of the transformed images.

In this work, the evaluation of our algorithms is done in two ways. First, we perform a combination of quantitative and qualitative analyses to determine the robustness of the normalization. For the quantitative analysis, we manually annotate a breast mask for 20 images both before normalization and after normalization for GE to Siemens as well as Siemens to GE, respectively. We then compute the Dice coefficient between these annotations. A higher Dice coefficient suggests that the normalization successfully preserved breast shape, while a lower value indicates distortion in breast shape. To evaluate the preservation of dense tissue, we perform qualitative analysis through visual observation.

Secondly, we evaluate the intensity normalization by manually annotating the dense tissue (10 cases) and subsequently computing the mean intensity value before and after normalization. The expectation is that while the mean intensities of dense tissue differ significantly between GE and Siemens before the normalization, they should be similar after the normalization.

\begin{table}[pos=b]
\begin{center}
      \caption{Dice coefficients between breast mask before and after normalization obtained on validation data.}
      \label{Tab:FOV_Val}
    \begin{tabular}{lllll} \toprule
          & \multicolumn{2}{c}{GE to SE} & \multicolumn{2}{c}{SE to GE} \\ \cline{2-5}
          FOV & Mean          & Std         & Mean          & Std          \\ \midrule
    $1\times 1$   & -             & -            & -             & -            \\
    $34\times 34$ & \textbf{0.9762}       & \textbf{{}0.0091}      & \textbf{0.9794}       & \textbf{0.0070}  \\
    $45\times 45$ & 0.9236       & 0.0164      & 0.9310       & 0.0177      \\
    $70\times 70$ & 0.9138       & 0.0577      & 0.9021       & 0.0443\\ \bottomrule
    \end{tabular}
    
\end{center}
\end{table}

\begin{figure} [!b]
\begin{center}
\begin{tabular}{c}
\includegraphics[height=6.25cm]{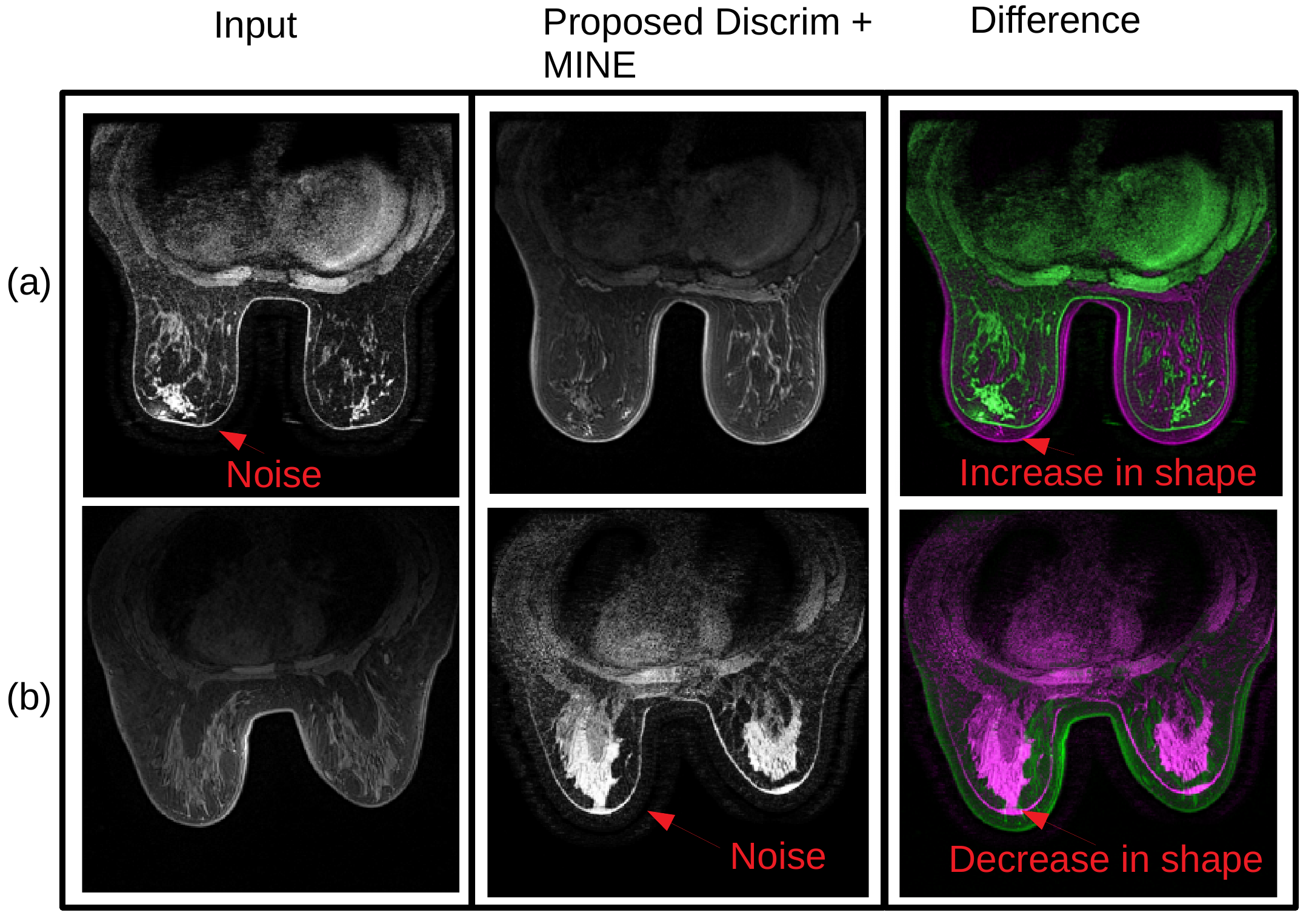}
\end{tabular}
\end{center}
\caption{Effect of mutual information loss (a) GE Healthcare to Siemens (b) Siemens to GE Healthcare. Difference between the input and normalized image is shown using a composite image, where magenta shows negative value and green shows the positive value. }
\label{fig:MI_Shape}
\end{figure}

\section{Results and Discussion}
\label{Sec:Result}

The result of the proposed MRI normalization using CycleGAN is presented in Fig.~\ref{fig:Qual_Result}. Qualitatively, it can be observed that the standard CycleGAN model is unable to preserve the shape of the breast and dense tissue. Our proposed modified discriminator framework performed the best out of all explored algorithms.

A surprising result visible in Fig.~\ref{fig:Qual_Result} is that the introduction of mutual information loss is unable to preserve the shape of the breast. After further analysis, we determine that the noise pattern in the GE images is the primary cause of this failure. Specifically, the mutual information neural estimator (MINE) network tries to maximize the mutual information by matching the shape of the breast to the noisy \textquotedblleft halo\textquotedblright~around the breast, and in doing so, actually increases the size of the breast. Similarly, for the Siemens to GE normalization, it maximizes the mutual information by decreasing the shape of the breast. This is illustrated in Fig.~\ref{fig:MI_Shape}.

\begin{figure}[!t]
\centering
	\includegraphics[width=.9\linewidth]{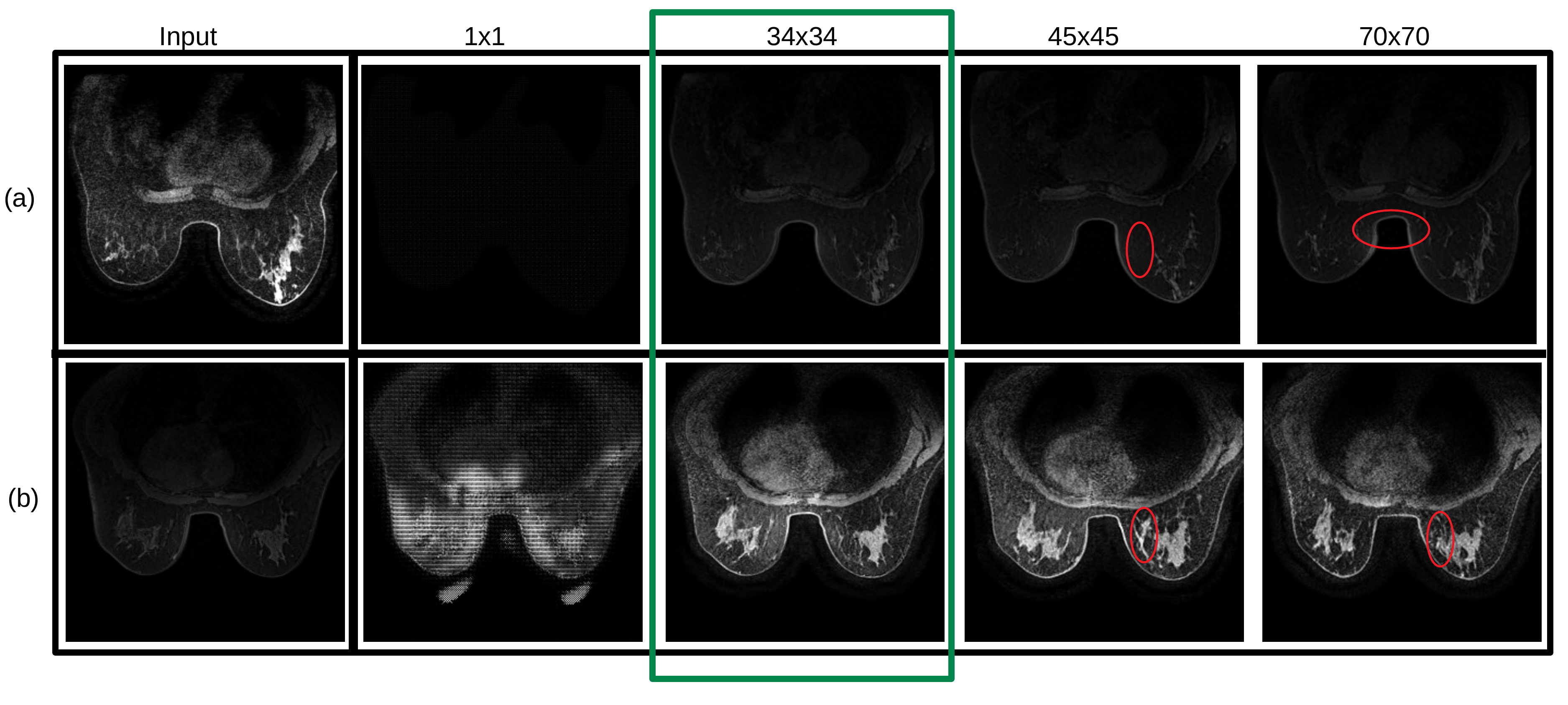}
\caption{Experiment with the field of view (FOV) in the discriminator architecture (a) GE Healthcare to Siemens (b) Siemens to GE Healthcare. Red ellipse shows the change in breast shape and dense tissue structure as compared to input.}
\label{fig:FOV}
\end{figure}

We propose a modified CycleGAN framework that involves altering the discriminator architecture in order to put more stress on features pertaining to breast tissue. We experiment with various FOV in the discriminator architecture and present the effects that these changes have on performance in Fig.~\ref{fig:FOV}. It can be observed that the $70\times 70$ FOV frequently modifies the dense tissues of the breast. It also modifies the shape of the breast, which is apparent from the lower Dice coefficients (Table~\ref{Tab:FOV_Val}). A $1\times1$ FOV, i.e. PixelGAN, has no effect on spatial statistics and is thus unable to learn the mapping between the noise distributions of the two domains. Additionally, the normalized images look extremely pixelated and exhibit a checkerboard pattern. The performance of a $45\times 45$ FOV is comparatively better than the $70\times 70$ FOV in terms of preserving both the breast shape as well as the dense tissue structures. However, visual inspection leads us to conclude that the  $34\times34$ FOV discriminator preserves the dense tissue better and produces sharper images compared to the $45\times 45$ FOV. The Dice coefficients confirm that the $34\times34$ FOV is able to preserve the shape of the breast as well. This improved performance, along with its lower number of parameters, lead us to select the $34\times 34$ FOV discriminator architecture.

Quantitative results are presented in Table~\ref{Tab:Dice_Coff_Final_4_1}. It can be observed that during GE to Siemens normalization, the Dice coefficient of the breast masks is the highest for the CycleGAN framework obtained by modifying the discriminator architecture. It is also apparent from Table~\ref{Tab:Dice_Coff_Final_4_1} that applying the mutual information loss to the proposed discriminator causes a reduction in the Dice coefficient value ($0.9801\rightarrow 0.9082$) due to a decrease in the shape of the breast. However, the standard CycleGAN model and its variant with mutual information both have comparable Dice coefficients. This can be explained by both methods' inability to preserve the breast shape.
Similar observations can be made for the normalization between Siemens to GE. This confirms that the modified architecture with the field of view of $34\times 34$ results in superior performance.

\begin{table}[pos=b]
\begin{center}
\caption{Quantitative results: Dice coefficients between breast mask before and after normalization on test data.}
        \label{Tab:Dice_Coff_Final_4_1}
        \begin{tabular}{lllll} \toprule
                              & \multicolumn{2}{c}{GE to SE} & \multicolumn{2}{c}{SE to GE} \\ \cline{2-5}
                         Models     & Mean          & Std          & Mean          & Std          \\ \midrule
        Std. CycleGAN          & 0.8913       &	0.0941      &	0.9089      & 	0.0471     \\
        Std. CycleGAN + MINE       & 0.8976       & 0.0510      & 0.8949       & 0.0391      \\
        Proposed Discrim        & \textbf{0.9801}       & \textbf{0.0061}      & \textbf{0.9813}       &  \textbf{0.0049}     \\
        Proposed Discrim + MINE & 0.9082       & 0.0714      & 0.8912       & 0.0706 \\  \bottomrule
        \end{tabular}

\end{center}  
\end{table}

\begin{figure} 
\centering
\includegraphics[width=1.0\linewidth]{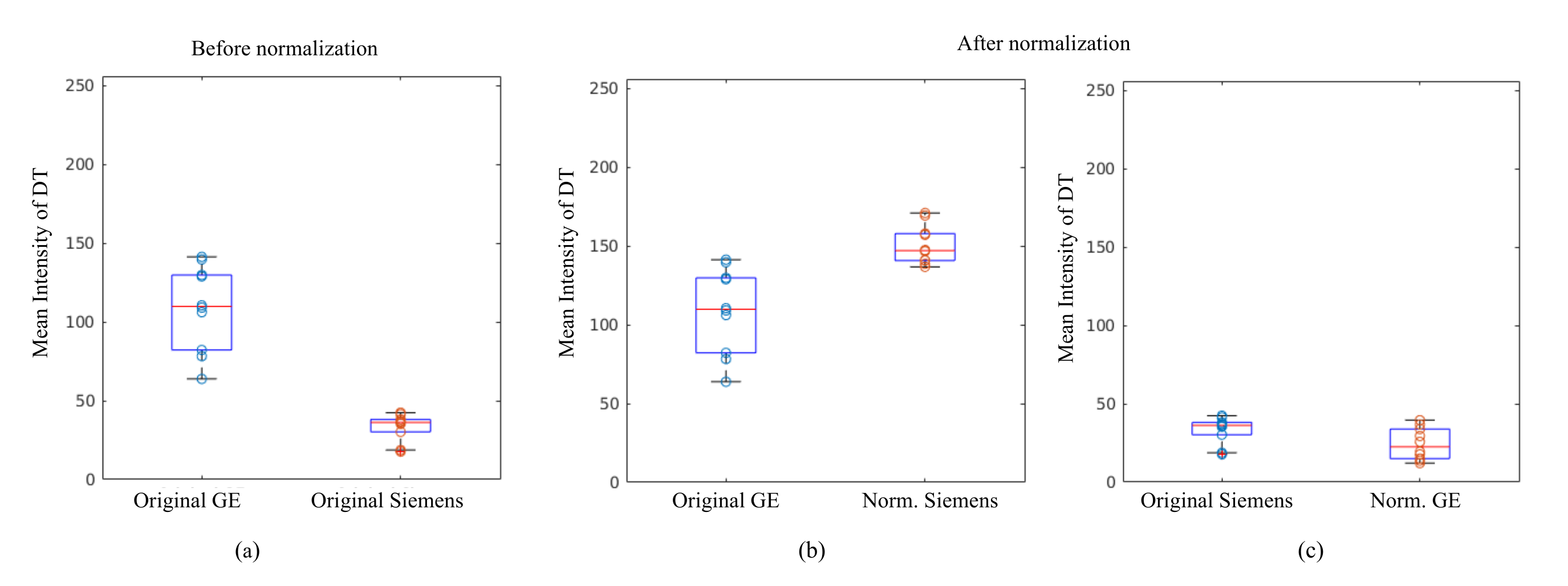}
\caption{Mean intensity value distribution of dense tissues (DT) in (a) original GE and original Siemens (b) original GE and normalized Siemens (c) and original Siemens and normalized GE.}
  \label{fig:Int}
\end{figure}

In summary, from a qualitative point of view, the standard CycleGAN along with mutual information leads to the worst result (See Fig.~\ref{fig:Qual_Result}) This is also reflected in the quantitative results, where it achieves almost the minimum dice coefficient score. On the other hand, the proposed modified CycleGAN framework obtained by altering the discriminator architecture is able to consistently preserve the dense tissue as well as the breast shape. These observations also align with the quantitative results on test data presented in Table~\ref{Tab:Dice_Coff_Final_4_1}.

To evaluate the intensity transformation, we manually annotate the dense tissue in 10 cases and then measure the mean intensity of these annotated regions both before normalization and after normalization. The result is presented in Fig~\ref{fig:Int} where Fig.~\ref{fig:Int}(a) illustrates the mean intensity distribution of the dense tissue in GE and Siemens before the normalization. It can be observed from Fig.~\ref{fig:Int}(b) that the mean intensity distribution of the original GE is comparable to the normalized Siemens. A similar observation can also be made from Fig.~\ref{fig:Int}(c) for the original Siemens and normalized GE. This demonstrates that the proposed method is able to successfully adjust the intensity of the image as it pertains to dense tissue. It should also be noted that along with intensity adjustment, the proposed method learns to map the noise \textquotedblleft halo\textquotedblright~around the breast, which is a crucial aspect of vendor normalization. The proposed vendor normalization method will thus potentially increase the robustness of downstream models that do not have access to adequate training data from multiple vendors by synthesizing larger and richer datasets, which will mitigate issues related to class imbalance.

\section{Conclusions}
\label{Sec:Conclusion}
In this article, we have shown that a fully convolutional neural network can be successfully trained to learn a bidirectional mapping and perform normalization between DCE-MRI images generated from different scanners (GE Healthcare \& Siemens). In contrast to previous works, our proposed method not only performs intensity normalization but also learns the noise distribution pattern.

Our evaluation shows that when the standard CycleGAN is applied to this task, it matches the desired intensity of images but struggles with the shape of the breast and dense tissue. This is caused by the limited constraint on the images generated by the GANs and in turn, liberty that it takes to freely generate breast images. In response to this, we propose two solutions. The first one is to incorporate mutual information into the loss function. Our rationale is that this modification will ensure that the structure of the breast is maintained between the input and the output of the generator. This first solution fails to solve the problem due to a very specific characteristic of the data, which is the noise \textquotedblleft halo\textquotedblright~around the breast. Incorporating mutual information into a CycleGAN is not a trivial task and we believe that the method of doing so proposed in this paper will be helpful for other similar tasks in medical imaging and beyond. The second solution to the problem of maintaining the structure of the breast that we propose in this paper is a modification to the discriminator. This solution proves to be highly successful for this task as verified by our experiments.

Our study has some limitations. One limitation of this work is that it provides the capability of translation using {2D} images only. While some effort in network design and parameter optimization is certainly needed, the proposed methods naturally lend themselves to 3D MR volumes. Another limitation is that our dataset consists of only two vendors and a relatively limited number of patients. While we still believe that the dataset used in this study represents the real-life problem faced in analyses of breast MRIs, further studies are needed to show that the proposed method generalizes beyond the data presented here. Finally, while we were able to demonstrate that our method results in no or minimal changes to the dense tissue structure, additional validation of the applicability for specific applications should be a topic of future studies. For example, while for some clinical applications, no changes in the breast tissue structure are acceptable, radiomics or deep learning applications are likely to be robust to some changes of this type. Data generated using our method could additionally be used to augment training data and improve deep learning model generalization. A deep learning model trained with augmented data from the various scanner will enable model generalization to real-world datasets with moderately different characteristics.

In summary, we propose a framework for normalization of breast MRIs based on CycleGAN. We also propose a few technical innovations that overcome various challenges that we experienced while applying CycleGAN framework to our task of breast MRI normalization. While the framework has only been tested using breast MRIs, it naturally lends itself to other medical imaging tasks where no paired data is available.

%%%%%%%%%%%%%%%%%%%%%%%%%%%%%%%%%%%%%%%%%%%%%%%%%%%%%%%%%%%%%%%%%%%%%%%%%%%%%%%%%%%%%%%%%%%%%%%%%%%%%%

\section*{Appendix}
\appendix
\section{ Architectures for discriminator}

\setcounter{table}{0}
\renewcommand{\thetable}{A.\arabic{table}}

%\label{Sec:Appendix}
\label{appendix:Disc_Arc}
Discriminator architectures with various field of view is presented in this section. Each model uses a convolution after the last layer to produce a 1-D output of size $m\times m$. Instance Norm layer was not applied to first layer in each of the architecture. The slope for LeakyReLU was 0.2.

 \begin{table}[pos=h]
  \label{Tab:70_70}
      \caption{Discriminator architecture ($70\times 70$)}
\begin{tabular}{llllll} \hline
Layer       & \begin{tabular}[c]{@{}l@{}}Input\\ Channel\end{tabular} & \begin{tabular}[c]{@{}l@{}}Output\\ Channel\end{tabular} & \begin{tabular}[c]{@{}l@{}}Filter\\Size (k)\end{tabular} & \begin{tabular}[c]{@{}l@{}}Stride\\ (S)\end{tabular} & Activation \\ \hline 
Convolution & 1                                                       & 64                                                       & $4\times 4$                                                       & 2                                                    & Leaky ReLU \\
Convolution & 64                                                      & 128                                                      & $4\times 4$                                                       & 2                                                    & Leaky ReLU \\
Convolution & 128                                                      & 256                                                      & $4\times 4$                                                     & 2                                                    & Leaky ReLU \\
Convolution & 256                                                     & 512                                                     & $4\times 4 $                                                      & 1                                                    & Leaky ReLU \\
Convolution & 512                                                     & 1                                                        & $4\times 4$                                                       & 1                                                    &   -      \\  \hline
\end{tabular} 
\end{table}

  \begin{table}[pos=h]
  \label{Tab:45_45}
     \caption{Discriminator architecture ($45\times 45$)}
\begin{tabular}{llllll} \hline
Layer       & \begin{tabular}[c]{@{}l@{}}Input\\ Channel\end{tabular} & \begin{tabular}[c]{@{}l@{}}Output\\ Channel\end{tabular} & \begin{tabular}[c]{@{}l@{}}Filter\\Size (k)\end{tabular} & \begin{tabular}[c]{@{}l@{}}Stride\\ (S)\end{tabular} & Activation \\ \hline 
Convolution & 1                                                       & 64                                                       & $5\times 5$                                                       & 2                                                    & Leaky ReLU \\
Convolution & 64                                                      & 128                                                      & $5\times 5$                                                       & 2                                                    & Leaky ReLU \\
Convolution & 128                                                     & 256                                                      & $5\times 5$                                                       & 1                                                    & Leaky ReLU \\
Convolution & 256                                                     & 1                                                        & $5\times 5$                                                        & 1                                                    &   - \\ \hline  
\end{tabular} 
\end{table}

 \begin{table}[pos=h]
  \label{Tab:34_34}
  \caption{Discriminator architecture ($34\times 34$)}
\begin{tabular}{llllll} \hline
Layer       & \begin{tabular}[c]{@{}l@{}}Input\\ Channel\end{tabular} & \begin{tabular}[c]{@{}l@{}}Output\\ Channel\end{tabular} & \begin{tabular}[c]{@{}l@{}}Filter\\Size (k)\end{tabular} & \begin{tabular}[c]{@{}l@{}}Stride\\ (S)\end{tabular} & Activation \\ \hline 
Convolution & 1                                                       & 64                                                       & $4\times 4$                                                       & 2                                                    & Leaky ReLU \\
Convolution & 64                                                      & 128                                                      & $4\times 4$                                                       & 2                                                    & Leaky ReLU \\
Convolution & 128                                                     & 256                                                      & $4\times 4$                                                       & 1                                                    & Leaky ReLU \\
Convolution & 256                                                     & 1                                                        & $4\times 4$                                                        & 1                                                    &   -   \\ \hline   
\end{tabular} 
\end{table}

\begin{table}[pos=h]
\label{Tab:pixel}
\caption{Discriminator architecture (PixelGAN)}
\begin{tabular}{llllll} \hline
Layer       & \begin{tabular}[c]{@{}l@{}}Input\\ Channel\end{tabular} & \begin{tabular}[c]{@{}l@{}}Output\\ Channel\end{tabular} & \begin{tabular}[c]{@{}l@{}}Filter\\Size (k)\end{tabular} & \begin{tabular}[c]{@{}l@{}}Stride\\ (S)\end{tabular} & Activation \\ \hline 
Convolution & 1                                                       & 64                                                       & $1\times 1$                                                       & 1                                                    & Leaky ReLU \\
Convolution & 64                                                      & 128                                                      & $1\times 1$                                                       & 1                                                    & Leaky ReLU \\

Convolution & 128                                                     & 1                                                      & $1\times 1$                                                      & 1                                                    & - \\  \hline
    
\end{tabular}
\end{table} 
 
\begin{table}[]
\centering
\caption{Hyperparameter ($\lambda_{cyc}$): Dice coefficients between breast mask before and after normalization obtained on validation data}
\label{Tab:Hyper_Cyclic_Loss}
\begin{tabular}{lllll} \hline
      & \multicolumn{2}{c}{GE to SE} & \multicolumn{2}{c}{SE to GE} \\ \cline{2-5}
      & Mean          & Std         & Mean          & Std          \\  \hline
 
$\lambda_{cyc}$ = 10   & 0.85660 & \textbf{0.04784} & 0.87012 & 0.03411 \\ 
$\lambda_{cyc}$  = 7.5  & 0.85760 & 0.07617 & 0.88693 & 0.03765 \\ 
$\boldsymbol{\lambda_{cyc}}$ \textbf{= 5.0} & \textbf{0.91381} & 0.05770 & \textbf{0.90209} & 0.04426 \\
$\lambda_{cyc}$ = 2.5 & 0.90243 & 0.08136 & 0.90577 & \textbf{0.03307} \\ \hline
\end{tabular}
\end{table}
 
\begin{table}[]
\centering
\caption{Hyperparameter ($\lambda_{mut}$): Dice coefficients between breast mask before and after normalization obtained on validation data}
\label{Tab:Hyper_Mut_Loss}
\begin{tabular}{cllll} \hline 
      & \multicolumn{2}{c}{GE to SE} & \multicolumn{2}{c}{SE to GE} \\ \cline{2-5}
      & Mean          & Std         & Mean          & Std          \\ \hline
$\lambda_{mut}$ = 0.01              & 0.87322                  & 0.05853                          & 0.90161                  & \textbf{0.03569}        \\
$\lambda_{mut}$ = 0.10              & 0.91552                  & \textbf{0.03438}                 & 0.89473                  & 0.03570                 \\
$\lambda_{mut}$ = 0.25              & 0.91112                  & 0.04433                          & 0.89564                  & 0.05675                 \\
$\boldsymbol{\lambda_{mut}}$ \textbf{= 0.50}     & \textbf{0.91801}         & 0.04055                          & \textbf{0.92319}         & 0.03903                 \\
$\lambda_{mut}$ = 1.0               & 0.88428                  & 0.04350                          & 0.91006                  & 0.03929                \\ \hline
\end{tabular}
\end{table}

\section*{Acknowledgments}
The authors would like to thank Mr. Mateusz Buda for the insightful discussions.

% \section*{References}

% Please ensure that every reference cited in the text is also present in
% the reference list (and vice versa).

% \section*{\itshape Reference style}

% Text: All citations in the text should refer to:
% \begin{enumerate}
% \item Single author: the author's name (without initials, unless there
% is ambiguity) and the year of publication;
% \item Two authors: both authors' names and the year of publication;
% \item Three or more authors: first author's name followed by `et al.'
% and the year of publication.
% \end{enumerate}
% Citations may be made directly (or parenthetically). Groups of
% references should be listed first alphabetically, then chronologically.

%%Harvard
%  \bibliographystyle{model2-names.bst}
% % \biboptions{authoryear}
% %\bibliography{refs}
% %\bibliography{bare_jrnl}
% % % \section*{Supplementary Material}
% \bibliographystyle{model2-names.bst}
% \biboptions{authoryear}
% % %\bibliography{refs}
% \bibliography{bare_jrnl}
% \section*{Supplementary Material}
% Supplementary material that may be helpful in the review process should
% be prepared and provided as a separate electronic file. That file can
% then be transformed into PDF format and submitted along with the
% manuscript and graphic files to the appropriate editorial office.

%%%%%%%%%%%%%%%%%%%%%%%%%%%%%%%%%%%%%%%%%%%%%%%
%%%%%%%%%%%%%%%%%%%%%%%%%%%%%%%%%%%%%
%%%%%%%%%%%%%%%%%%%%%%%%%%

%\bibliographystyle{cas-model2-names}

\bibliographystyle{unsrtnat}

% % Loading bibliography database
\bibliography{cas-refs}

% %\vskip3pt

% % \bio{}
% % Gourav Modanwal is currently working as a Research Associate at Duke University, Durham, NC, USA. He has received his Ph.D. degree from Indian Institute of Technology (BHU), Varanasi, India in 2019. His research interests include Human-Computer Interaction, Computer Vision \& Pattern Recognition, and it's application to the medical domain.
% % \endbio

% % \bio{figs/pic1}
% % Author biography with author photo.
% % Author biography. Author biography. Author biography.
% % Author biography. Author biography. Author biography.
% % Author biography. Author biography. Author biography.
% % Author biography. Author biography. Author biography.
% % Author biography. Author biography. Author biography.
% % Author biography. Author biography. Author biography.
% % Author biography. Author biography. Author biography.
% % Author biography. Author biography. Author biography.
% % Author biography. Author biography. Author biography.
% % \endbio

% % \bio{figs/pic1}
% % Author biography with author photo.
% % Author biography. Author biography. Author biography.
% % Author biography. Author biography. Author biography.
% % Author biography. Author biography. Author biography.
% % Author biography. Author biography. Author biography.
% % \endbio

\end{document}